\documentclass[
reprint,
superscriptaddress,
nofootinbib,
amsmath,amssymb,
aps,
]{revtex4-2}

\usepackage{graphicx}
\usepackage{dcolumn}
\usepackage{bm}
\usepackage[utf8]{inputenc}
\usepackage{physics}
\usepackage{xcolor}
\usepackage{upgreek}
\usepackage{footnote}

\begin{document}

\title{Towards integrated sensors for optimized OCT with undetected photons}

\author{Franz Roeder}
\email{franz.roeder@uni-paderborn.de}
\affiliation{Paderborn University, Institute for Photonic Quantum Systems (PhoQS), Warburger Str. 100, 33098 Paderborn, Germany}
\affiliation{Paderborn University, Integrated Quantum Optics, Warburger Str. 100, 33098 Paderborn, Germany}
\author{René Pollmann}
\affiliation{Paderborn University, Institute for Photonic Quantum Systems (PhoQS), Warburger Str. 100, 33098 Paderborn, Germany}
\affiliation{Paderborn University, Integrated Quantum Optics, Warburger Str. 100, 33098 Paderborn, Germany}
\author{Viktor Quiring}
\affiliation{Paderborn University, Institute for Photonic Quantum Systems (PhoQS), Warburger Str. 100, 33098 Paderborn, Germany}
\affiliation{Paderborn University, Integrated Quantum Optics, Warburger Str. 100, 33098 Paderborn, Germany}
\author{Christof Eigner}
\affiliation{Paderborn University, Institute for Photonic Quantum Systems (PhoQS), Warburger Str. 100, 33098 Paderborn, Germany}
\author{Benjamin Brecht}
\affiliation{Paderborn University, Institute for Photonic Quantum Systems (PhoQS), Warburger Str. 100, 33098 Paderborn, Germany}
\affiliation{Paderborn University, Integrated Quantum Optics, Warburger Str. 100, 33098 Paderborn, Germany}
\author{Christine Silberhorn}
\affiliation{Paderborn University, Institute for Photonic Quantum Systems (PhoQS), Warburger Str. 100, 33098 Paderborn, Germany}
\affiliation{Paderborn University, Integrated Quantum Optics, Warburger Str. 100, 33098 Paderborn, Germany}

\begin{abstract}
The development of practical sensors for optical coherence tomography (OCT) with undetected photons requires miniaturization via integration. To be practical, these sensors must exhibit a large spectral bandwidth and a high brightness, which are linked to a high axial resolution and a sufficient signal to noise ratio, respectively. Here, we combine these requirements in a scheme for OCT-measurements with undetected photons based on nonlinear Ti:LiNbO$_3$ waveguides. We investigate the performance benchmarks of the commonly used SU(1,1) scheme in comparison to an induced coherence scheme and find that the latter is actually better suited when implementing measurements with undetected photons in integrated systems. In both schemes, we perform pump gain optimization and OCT measurements with undetected photons with an axial resolution as low as $28\,\mathrm{\upmu m}$.
\end{abstract}

\maketitle

\section{Introduction}

Classical ultra-fast spectroscopy employs spectrally broadband light to measure spectral and spatial features of an object under test with high resolution. Typical systems rely on interferometric schemes where the pulse is split into a probe and a reference path. After interaction of the probe pulse with the sample, spectral or temporal interference of probe and reference is recorded. This allows one to obtain spatial depth information (via optical coherence tomography \cite{Leitgeb2003}) or spectral information (via Fourier transform spectroscopy \cite{Griffiths2007}) of a sample under test. Naturally, probing and detection occur at the same wavelength in these systems. This leads to challenges for the detection when moving to wavelengths in the mid infra-red (MIR) spectral range and prohibits applications in the low light regime.\\
Lately the use of quantum nonlinear interferometers, based on non-degenerate photon pair sources, opened up new pathways to circumvent this problem and allow for techniques that are referred to as "measurements with undetected photons". In these schemes, a photon at a long wavelength which is in a technically hard to access spectral region, e.g., the MIR, probes the object. Meanwhile, the second photon at a short wavelength serves as a reference and can be detected by broadly available detectors, e.g., in the near IR. Due to information transfer from the long to the short wavelength photon, properties of the object under test can be extracted at this shorter wavelength. Consequently, many techniques from the field of ultra-fast spectroscopy have been realized in this scheme; examples include spectroscopy \cite{Lindner2021, Paterova2018, Kaufmann2022}, imaging \cite{Lemos2014, Toepfer2022, Kviatkovsky2020, Torres2024} or OCT \cite{Rojas-Santana2021, Valles2018} with undetected photons.\\

OCT generally comes in two operation modes: time domain (TD) and Fourier-domain (FD). In classical OCT-devices, FD-OCT is often favoured as it allows one to record a single-shot spectral interferogram and does not require moving parts. However, when operating with single photons, expensive spectrometers and long integration times are required. Therefore, TD-OCT becomes appealing at low light levels. Here, spectrally unresolved measurements can be carried out in a scanning approach with affordable single photon detectors.\\ 
Generally, quantum nonlinear interferometers come in two flavors which differ in the way in which the interference is realized. They are referred to as the induced coherence (IC) \cite{Wang1991} and SU(1,1) \cite{Yurke1986} scheme, respectively, and will be discussed in more detail later. One of the first realizations of measurements with undetected photons has been performed in a setup based on IC, first presented in 1991 \cite{Wang1991}. The first practical application demonstrated imaging with undetected photons with different colours \cite{Lemos2014}. Since then, the SU(1,1) configuration has gained more prominence due to the ease in alignment for experiments employing bulk nonlinear crystals. Recently, both setups have also been combined in order to design hybrid setups that allow for changes in the sensing modes \cite{Gemmell2024, Kim2024} or even been investigated in terms of the quantum optical properties in OCT \cite{Rojas-Santana2021, Valles2018, Machado2020, Kim2024}.\\
So far, most experimental realizations are based on bulk nonlinear crystals. For practical applications with widespread industrial use, however, integration is required. This comes with its own challenges: waveguides offer a long interaction length, which increases source brightness but generally limits the spectral bandwidth, unless careful dispersion engineering is employed \cite{Eckstein2011, Pollmann2024, Roeder2024_source}; furthermore, due to the different wavelengths involved in measurements with undetected photons, waveguides can support multiple spatial modes which limits the interference in one specific mode.\\

\begin{figure*}[t!]
	\centering
	\vspace{5mm}
	\includegraphics[width=0.85\linewidth]{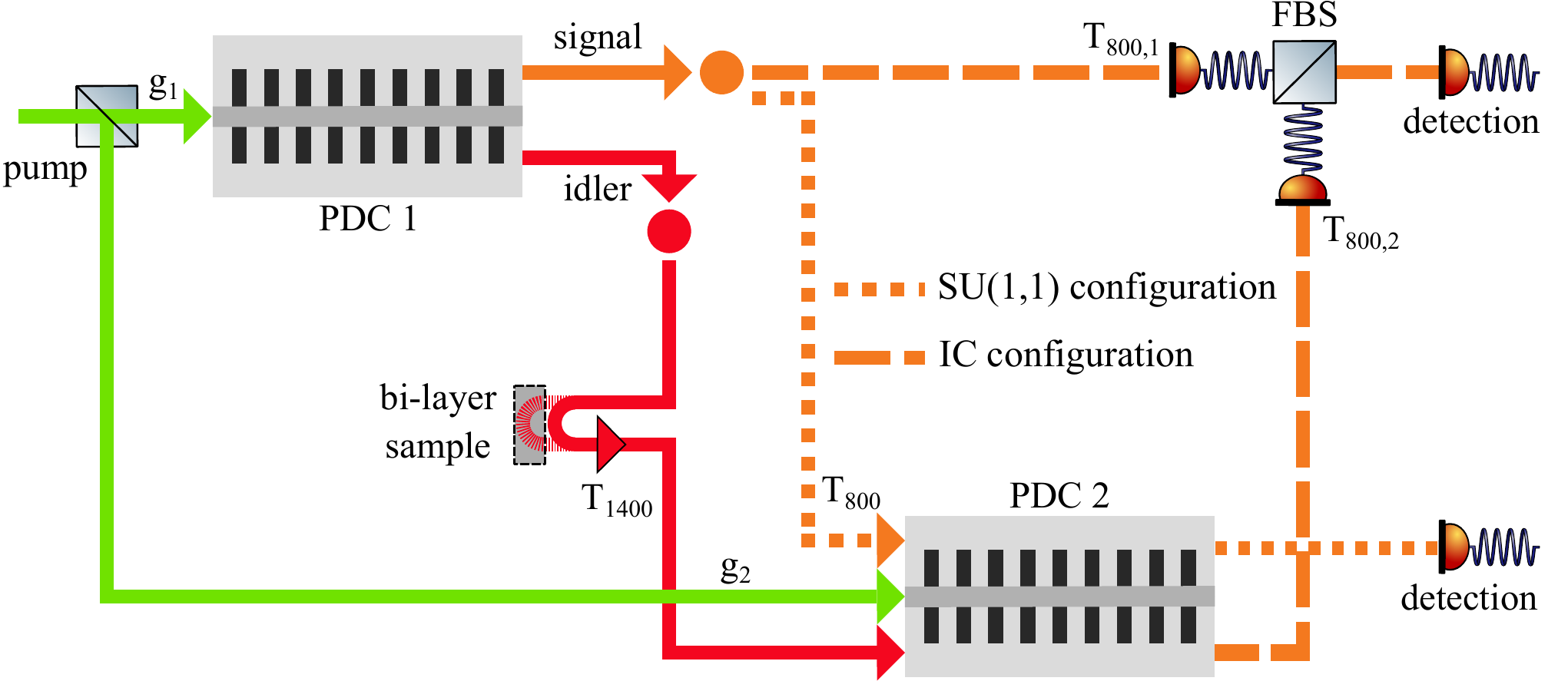}
	\caption{Concept of a nonlinear interferometer in the SU(1,1) geometry (dotted) and in the IC geometry (dashed). In both cases, the idler photon from the first PDC process interacts with a bi-layer sample. The signal  photon from the first process is overlapped with the one from the second process either in the second PDC source in the SU(1,1) case or on a fibre beam splitter (FBS) in the IC case.}
	\label{fig:concept}
\end{figure*}

In this work, we establish a detailed framework for the design and realization of integrated sensors for OCT with undetected photons. To this end, we demonstrate - in theory and experiment - how dispersion-engineered, broadband integrated photon pair sources can be optimally used for the development of a practical quantum sensor comparing both the SU(1,1) and the IC scheme. This comparison for an integrated platform adds to current developments towards interferometers operating in hybrid setups \cite{Gemmell2024, Kim2024}, highlighting that different geometries can have benefits for specific applications.\\
In our investigation, we first identify the optimal operation points of the pump gain ratio of both schemes. Subsequently, we find that the IC scheme does not only offer improved axial resolution when compared to the SU(1,1) scheme, but also simplifies system design and simultaneously improves robustness thanks to inherent single-spatial mode operation for the probe and reference photon. Furthermore, we benchmark the performance of both systems against each other in terms of the axial resolution, pump power consumption and axial scanning range of our system in the FD-OCT and TD-OCT operation mode. As a result, we identify TD-OCT in the IC scheme as the best candidate for the construction of an integrated quantum sensor.

\section{Results}\label{sec:results}

\subsection{Theoretical description}

Fig.\,\ref{fig:concept} shows a schematic realization of a nonlinear interferometer in the SU(1,1) and IC configuration. The interferometer is constructed from two identical, cascaded waveguide-based parametric down-conversion processes (PDC 1 and PDC 2) that are pumped with a common pump laser to ensure coherence. The generated long-wavelength photon (idler) from PDC 1 probes an optional target, in our case a bi-layer sample with partially reflective surfaces, before it is coupled into PDC 2. The two processes are made indistinguishable by carefully aligning the paths of the idler photons from PDC 1 and PDC 2 on top of each other, which leads to quantum interference between the photon pair generation probability amplitudes of the processes PDC 1 and PDC 2. Thereby, information is transferred from the long wavelength photon of PDC 1 onto the short wavelength photon (signal) of PDC 2. This information can then be uncovered upon interference of the signal photons from both processes.\\
In the IC configuration, the signal photons from PDC 1 and PDC 2 need to be interfered on an external beam splitter. In the experiment, the two signal fields are coupled to single mode fibres and interfered on an integrated fibre beam splitter. Interference is then detected at one output port of the beam splitter.\\
In the SU(1,1) configuration, the signal from PDC 1 is sent through PDC 2 together with the idler. Here, interference is detected directly at the output of PDC 2. \\

For our OCT measurements, we place the aforementioned bi-layer sample, which we define to have a layer separation $d$ and layer reflectivities $r_{1}$ and $r_2$, in the idler arm as depicted in Fig.\,\ref{fig:concept}. Here, $r_1, r_2$ are real numbers that are chosen such that the incoming light is reflected by either one or the other layer. The sample imprints a complex reflection coefficient $r(\Delta \omega)$ onto an incoming idler light field as 

\begin{equation}
	r(\Delta \omega) = r_1 + r_2 \exp(i \cdot \frac{2 n_g d  \Delta \omega}{c})
\end{equation}

assuming a constant group refractive index $n_g$ over the spectral range of the incident PDC light. This is justified as the difference in the group indices amounts to $0.5\%$ over the spectral bandwidth covered in the experiment. Furthermore, we consider a frequency detuning $\Delta \omega = \omega_s - \bar{\omega}_s= - (\omega_i - \bar{\omega}_i)$ around the signal and idler central frequencies $\bar{\omega}_{s,i}$ assuming a cw-pump laser.\\
With this object under test, the spectral interferogram at the signal wavelength can be written as

\begin{equation}
	\begin{aligned}
		S(\Delta \omega) \propto &\abs{f_{PDC}(\Delta \omega)}^2 [1 + r_1^2\cdot \cos(\Phi_{int}(\Delta \omega)+\tau \Delta \omega) \\&+ r_2^2\cdot \cos(\Phi_{int}(\Delta \omega)+\tau \Delta \omega + 2 n_g d/c \Delta \omega)].
		\label{eq:OCT_spec}
	\end{aligned}
\end{equation}  

This interferogram is determined by the joint spectral amplitude (JSA) of a single PDC process $f_{PDC}(\Delta \omega)$, which describes the spectral distribution of the generated PDC light around the central frequencies, an optical path delay between the two processes $\tau$ and an internal phase within the interferometer $\Phi_{int}$. Both of these terms are discussed further in the Appendix.\\
The obtained spectral interference pattern exhibits a beating between two frequencies. Upon Fourier transformation, we observe two distinct peaks with a width that is determined by the spectral width of the bi-photons contained in $f_{PDC}$ via the Wiener-Khinchin theorem \cite{Leitgeb2003, Vanselow2020} (see Appendix). These peaks can be shifted in order to position them away from the zero component by adjusting the optical path delay $\tau$ introduced in the idler arm. The distance of the sample layers with separation $2 n_g d/c$ as the round-trip optical path delay is given by the peak separation in this Fourier transform.\\
Similarly, the temporal interferogram can be obtained by photon counting at the signal output while varying the temporal delay $\tau$. This measurement is spectrally un-resolving and therefore equivalent to an integration over all spectral components. Thus, the observed signal count rate is given by

\begin{equation}
	C_s(\tau) = \int_{-\infty}^{\infty} \mathrm{d}\Delta \omega \, S(\Delta \omega, \tau). 
\end{equation}  

An interference pattern in the signal photon count rate can be observed every time that the time delay $\tau$ in the argument of the cosine terms in Eq.\,\ref{eq:OCT_spec} is opposite to the delay introduced by the separation between the sample layers. Thus, while scanning the reference delay, interference patterns are observed at delays that are separated by $2 n_g d/c$. The achievable axial resolution is again given by the bandwidth of the bi-photon source. \\
However, in practical systems with integrated PDC sources the internal phases such as second order dispersion lead to the generation of chirped bi-photons in the PDC state. This results in longer actual correlation times and an increased width of the temporal interferograms, as we discussed in \cite{Roeder2024}. Ultimately, this intrinsic dispersion during the generation of broadband photon pairs leads to a decreased axial resolution of the directly measured signals. The broadening of the signal can be numerically compensated for in post-processing as we discuss in the Supplemental Material.\\

Apart from the shape of the interference signal that defines the resolution and layer separation, a high signal-to-noise ratio (SNR) is necessary for a practical quantum sensor. When performing measurements with undetected photons, the interference visibility is an equivalent indicator of this SNR \cite{Aumann2019}. We achieve the highest visibility if the number of interfering photons from both processes is equal \cite{Santandrea2023}. Therefore, we consider different pump gains of the two PDC processes $g_{1,2}$ to maximize the spectral interference visibility. We link the experimentally accessible pump power $P$ to the gain parameter as $g \propto \sqrt{P}$. This allows to establish a link to the mean photon number of each PDC process given by $\langle n \rangle = \sinh^2 g \approx g^2$ \cite{Santandrea2023}. Here, we consider an operation in the single photon pair regime where the gains $g_{1,2} \ll 1$. This coincides with our assumption of generating at maximum one photon pair per joint spectral mode in either one of the two PDC processes.\\
In the following we adapt the calculation in \cite{Santandrea2023} which describes a SU(1,1) interferometer with losses between the two nonlinear processes and demonstrates how a high interference visibility can be restored by adjusting the pump gain ratios. Although this theory assumes a single spectral mode, we can still apply it to our spectrally multi-mode system by detecting the frequency resolved interferograms. Thereby, we evaluate the system in distinct, effective spectral bins. In the time resolved measurements, we expect a lower visibility due to second order dispersion. However, this visibility is still maximal for the extracted optimal pump gain ratios.\\
Differential pumping has been shown before in setups based on bulk nonlinear crystals \cite{Gemmell2023} and for pump gain regimes exceeding the low gain limit \cite{Kolobov2017}. Yet, when working with hybrid integrated setups, which combine waveguides and free space beam propagation, differential pumping becomes even more relevant for counteracting losses from coupling out and into the waveguide.\\ 
Specifically, the visibility at the signal wavelength around $800\,\mathrm{nm}$ in our measurements with undetected photons in the SU(1,1) scheme is given by \cite{Santandrea2023}:

\begin{equation}
		V_{800,SU(1,1)}=\frac{2 \sqrt{T_{800}T_{1400}}g_1 g_2}{T_{800}g_1^2+g_2^2}
\label{eq:vis_800_SU11}
\end{equation}

The relevant transmission and gain values are annotated in Fig.\,\ref{fig:concept}. We can observe that the maximum visibility is bound by the transmission in the idler arm $T_{1400}$ and is reached at a pump gain ratio $g_1^2/g_2^2 = 1/T_{800}$ which is given by the transmission in the signal arm $T_{800}$. For a guided-wave PDC source, these transmission values are reduced due to the coupling to the second source. Thus a second waveguide source typically requires weaker pumping than the first one.\\
So far, differential pumping has only been investigated in the SU(1,1) configuration. However, we also want to study these effects in the IC configuration. We therefore expand on the above treatment and now derive an expression for the visibility in the IC case (see Supplemental Material). In this configuration, the two signal fields from both PDC processes are now coupled into two single mode fibres for interference on a fibre beam splitter. Such a beam splitter ensures perfect spatial overlap and high-quality interference. Thus, two transmission parameters that represent the fibre coupling efficiencies $T_{800,1}, T_{800,2}$ for the signal fields from the first and the second process have to be considered in the calculation of the interference visibility:

\begin{equation}
		V_{800,IC}=\frac{2 \sqrt{T_{800,1}T_{800,2}T_{1400}}g_1 g_2}{T_{800,1}g_1^2+T_{800,2}g_2^2}.
\end{equation}

It can be seen that the visibility is still bound by the idler transmission $T_{1400}$. Meanwhile, the optimal pump gain ratio $g_1^2/g_2^2 = T_{800,2}/T_{800,1}$ is now given by the ratio of the fibre coupling efficiencies from the first and second source. These efficiencies can be optimized - without interacting with the waveguide source at the core of the interferometer - via mode-matching for the fibre couplings and can reach values close to 1. Furthermore, the optimal pump gain is now determined only by the ratio of the fibre coupling efficiencies. This allows one to optimize the pump gain ratio, e.g., by attenuating one signal beam, without having to adjust the pump powers.\\
An additional technical advantage of the IC configuration is the elimination of the back-coupling of the signal field from the first process into a spatially multi-mode waveguide.

\begin{figure*}[t!]
	\centering
	\includegraphics[width=0.8\linewidth]{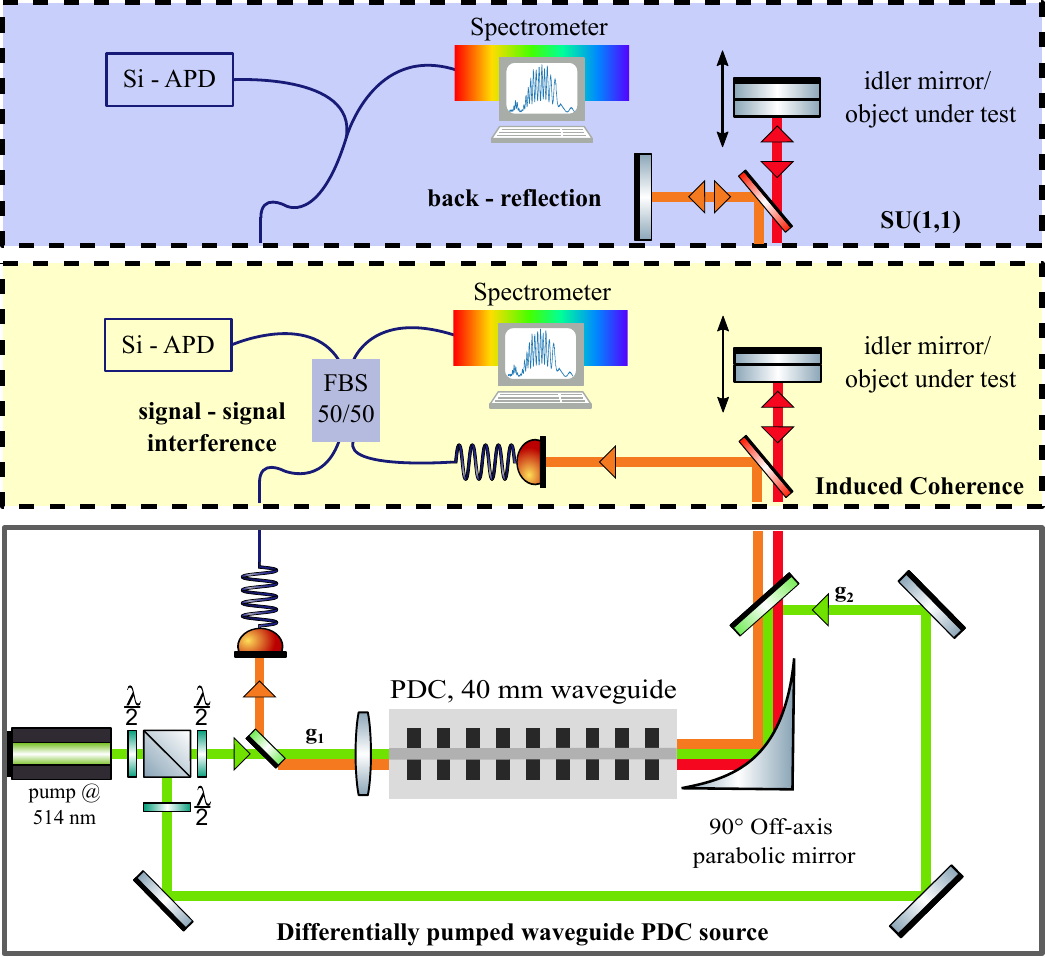}
	\caption{The experimental setup consists of a differentially pumped 40\,mm long Ti:LiNbO$_3$ waveguide as a PDC source. The generated photons from the first process are collimated by an off-axis parabolic mirror. Afterwards, either both photons are coupled back into the waveguide (SU(1,1) scheme) or the signal is routed around the waveguide to a fibre beam splitter to interfere with the signal from the second process while only the idler is coupled back to the waveguide (IC scheme). In both cases, the idler field interacts with a bi-layer object under test and a relative delay between signal and idler can be tuned by a motorized stage in the idler arm. The temporal and spectral interferograms are detected on a Silicon-APD or spectrometer, respectively.}
	\label{fig:setup}
\end{figure*}

\subsection{Experimental setup}

The central element of the interferometer as shown in Fig.\,\ref{fig:setup} is a $40\,\mathrm{mm}$ long, periodically poled Ti:LiNbO$_3$ guided-wave PDC source with matching signal and idler group-velocities, thus reaching spectral bandwidths of more than $6\,\mathrm{THz}$. The shape and bandwidth of the emitted PDC spectrum from this source can be tuned by adjusting the waveguide temperature \cite{Pollmann2024}.\\
In the experimental setup, we pump our waveguide with a $514\,\mathrm{nm}$ cw-laser, generating PDC light at $830\,\mathrm{nm}$ and $1360\,\mathrm{nm}$ in signal and idler fields, respectively. The pump gain ratio between the two PDC processes is adjusted by introducing a half-wave plate and PBS in front of the waveguide to split the pump power and achieve gains $g_1$ in forward direction and $g_2$ in backwards direction. The pump polarization is adjusted in both arms to match the requirement for the type II phasematching. \\
The signal and idler photons, which are generated in the first pass of the pump light through the waveguide, are collimated by an off-axis parabolic mirror (OAP). We separate the signal and idler photons using a dichroic mirror.\\
As illustrated in Fig.\,\ref{fig:setup}, we realise two different schemes for our nonlinear interferometer: the SU(1,1) scheme and the IC scheme, which are marked by the blue and yellow boxes, respectively. In both cases, the longer wavelength idler interacts with a layered object under test before it is coupled back into the waveguide. The end mirror in the idler arm is moved with a motorized stage to introduce a delay between signal and idler, resulting in a linear spectral phase. The signal photons are either reflected back and coupled to the waveguide in the SU(1,1) scheme, or directly coupled to a single mode fibre in the IC scheme. We note here, that the signal field is spatially multi-mode and therefore critically sensitive to misalignment when coupled back to the waveguide in the SU(1,1) configuration. This technical challenge is overcome in the IC configuration by directly coupling to single mode fibres which avoids the possibility of coupling to higher order spatial modes.\\
After the second PDC process, the interference at the signal wavelength is observed by separating the interfering signal photons from the pump and idler and coupling them to a single mode fibre. In the SU(1,1) scheme, the interference occurs during the second pass through the nonlinear waveguide and can be observed directly in the signal field exiting the waveguide. For the IC scheme, the interference at the signal wavelength can be observed by coupling the photons from the first and second pass to a 50:50 fibre beam splitter. Afterwards, both spectral and temporal interference can be observed in both configurations.

\begin{figure*}[t!]
	\centering
	\includegraphics[width=\linewidth]{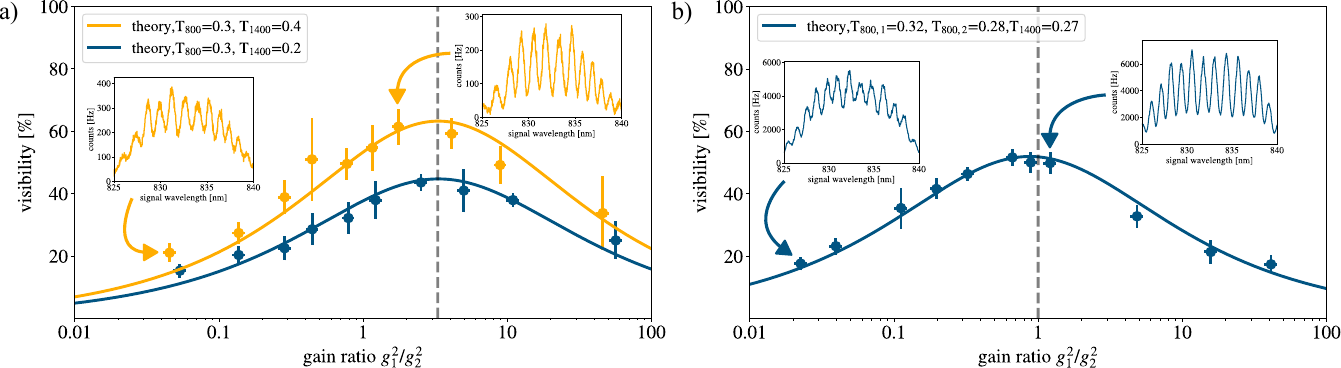}
	\caption{a) Gain optimization in the SU(1,1) geometry for two different transmission values in the idler arm at $1400\,\mathrm{nm}$ by using the reflection off a plain mirror or the front surface of a  partially reflective ND03 filter. The optimal pump gain ratio stays the same value around 3 for both cases and is highlighted by the dotted gray line. b) Gain optimization in the IC geometry with a plain mirror in the idler arm. The optimal gain ratio of $g_1^2/g_2^2 = 0.875$ is close to the highlighted value of $g_1^2/g_2^2= 1$ and given by the ratio of the fibre coupling efficiencies for the signal fields. The insets show recorded spectral interferograms at different pump gain ratios showcasing the increase in interference visibility.}
	\label{fig:gain_opt}
\end{figure*}

\subsection{Measurement results}

In this section, we provide the most important benchmarks of our system in the two different configurations. With that, we determine their suitability for the implementation of a practical quantum sensor. We concentrate on optimizing the interference visibility for a high SNR via pump gain optimization and discuss the achieved axial resolution in FD-OCT and TD-OCT. The impact of other parameters such as the signal losses, the system's sensitivity to low layer reflectivities and the effects of averaging data can be found in the Supplemental Material.\\
We start by optimizing the pump gain ratio in order to achieve the highest interference visibility in both experimental setups. Here, we measure the visibility of the spectral interferogram for different pump gain ratios and compare the obtained values to the theoretical expectations.\\
We present the results for the SU(1,1) scheme in Fig.\,\ref{fig:gain_opt} a). The visibility is measured first using a plain silver idler end mirror, followed by a measurement where we use the reflection off a reflective ND03 filter that was attached to the silver mirror. This leads to idler transmission values of $T_{1400} = 0.4$ and $T_{1400}=0.2$, respectively, which include losses due to coupling to the waveguide and from the partial reflectivity of the filter.\\
In both cases, we observe a maximum in the visibility for a pump gain ratio of $g_1^2/g_2^2 = 1/T_{800} \approx 3$, as expected from Eq.\,\ref{eq:vis_800_SU11}. This sets the optimal operation point of the SU(1,1) interferometer. It is important to note that this condition is independent of the idler transmission, hence every test object is probed with the optimal SNR.\\
Furthermore, we can determine the reflectivity of the object in the idler arm by comparing the two transmission values. Those transmission values are the only parameters used to match the experiment to the simulations and agree well with the expected coupling losses to the waveguide. We observe that, without further optimization of the data acquisition, our setup can correctly retrieve transmissions in the idler arm as low as $5\,\%$, which we show in Fig.\,S2 in the Supplemental Material. This highlights the capability to detect layers with low reflectivities in an object under test. Losses to the signal field, e.g., due to environmental impacts, can be mitigated by optimizing the pump gain ratio as investigated in Fig.\,S3 in the Supplemental Material.\\
In the IC scheme, the optimal pump gain ratio is not given by the coupling of the signal field to the waveguide, but by the ratio of the fibre coupling efficiencies $T_{800,1}$ and $T_{800,2}$. This behaviour can be nicely observed in our data shown in Fig.\,\ref{fig:gain_opt} b). Here, the theory curve reaches its maximum at $g_1^2/g_2^2 = T_{800,2}/T_{800,1}$ and is based on the experimentally determined fibre coupling efficiencies of $T_{800,1}=32\,\%$ for the signal from the first pass and $T_{800,2} = 28\,\%$ for the signal from the second pass.\\
The relevant fibre coupling efficiencies can be optimized without changing the alignment within the interferometer and can therefore reach values close to unity with adequate mode-matching. Here, the realised fibre coupling efficiencies of $32\,\%$ and $28\,\%$ lead to an optimal gain ratio of $0.875$, which is close to equal powers in both stages. Such a condition with equal pump powers is particularly useful as it allows for the realization of practical integrated sensors with a simplified pumping scheme: As we have already demonstrated in earlier work \cite{Roeder2024}, highly reflective coatings at the waveguide end facets can replace a second pump light beam path. The possibility to only pump the waveguide from one side, and still obtain the optimal pump gain ratio, makes this approach suitable for further integration of the setup into a practical sensor.\\

\begin{figure*}[t!]
	\centering
	\includegraphics[width=\linewidth]{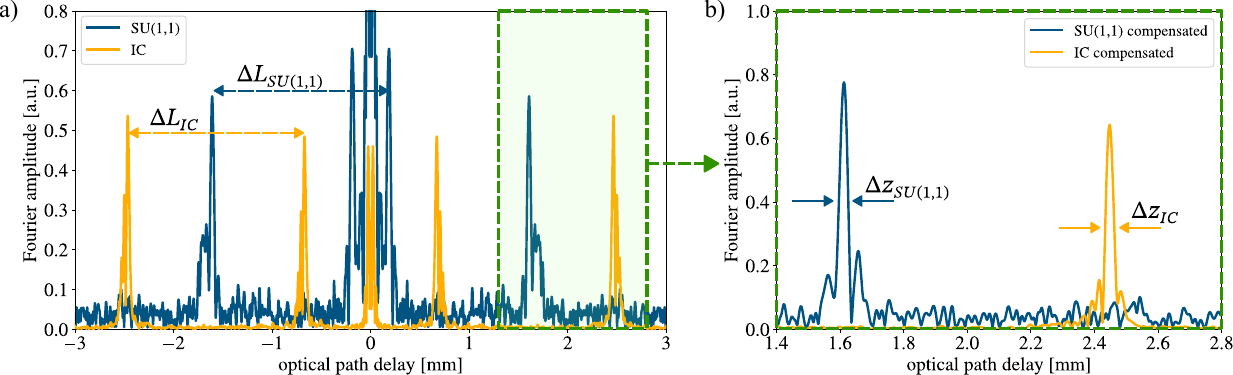}
	\caption{a) Fourier amplitude of the FD-OCT signal from the SU(1,1) and IC scheme used to obtain layer separations of ${\Delta L}_{SU(1,1)} = (1.790\pm 0.029) \,\mathrm{mm}$ and ${\Delta L}_{IC} = (1.790\pm 0.028) \,\mathrm{mm}$. b) Close-up on two peaks in the Fourier transformed signal of the FD-OCT for SU(1,1) and IC scheme revealing an axial resolution of $29.2\,\mathrm{\upmu m}$ and $28.1\,\mathrm{\upmu m}$ from the FWHM after dispersion compensation.} 
	\label{fig:FD_comp}
\end{figure*}

For demonstrating the capabilities of our experimental setup towards a practical sensor, we realise OCT-measurements of a bi-layer object under test consisting of a neutral density filter (ND03) attached to a plain silver mirror. This configuration results in two surfaces that reflect the incoming light in two equal portions and provides a single-trip optical path difference (OPD) of ${\Delta L}_{theo}= (1.50\pm 0.15)\,\mathrm{mm}$. Here, the ND filter's thickness and its variation are given by the manufacturer.\\
We record the FD-OCT and TD-OCT measurements of this sample for the IC and SU(1,1) experimental setup at their respective optimal pump gain ratios. In the measurements, our focus lies on determining the axial resolution of our measurement rather than the true OPD introduced by the sample which can be altered due to alignment and potential air-gaps.\\

Firstly, we perform FD-OCT in the SU(1,1) and IC setup with a total pump power of $5\,\mathrm{mW}$. After Fourier-transformation of the measured spectral interferograms, the layer separation can be extracted from the distance between the peaks in Fig.\,\ref{fig:FD_comp} a) which shows the amplitude of the Fourier-transform on the y-axis while the x-axis displays the single trip OPD. The measured raw spectral interferograms, which were used to extract the FD-OCT signal, are shown in Fig.\,S4 in the Supplemental Material.\\
The measurements in both experimental schemes have been performed with different delays introduced by the idler mirror. This leads to different shifts of the peak positions in both setups, but does not change their distance. We can extract a layer separation with ${\Delta L}_{SU(1,1)} = (1.790\pm 0.029) \,\mathrm{mm}$ and ${\Delta L}_{IC} = (1.790\pm 0.028) \,\mathrm{mm}$. These values for the OPD agree with each other and are larger than the expected OPD, which can be attributed to slight misalignment and air-gaps.\\
The measurements in the SU(1,1) and IC configuration show a similar axial resolution of $\Delta z_{FD,SU(1,1)} = 29.2\,\upmu\mathrm{m}$ for the SU(1,1) geometry and $\Delta z_{FD,IC} = 28.1\,\upmu\mathrm{m}$ for the IC geometry, which we retrieve from the FWHM peak width in the Fourier domain. The peaks from the sample's front layer, which have been used to extract these values, are depicted in a close-up of the Fourier transform in Fig.\,\ref{fig:FD_comp} b). Numerical dispersion compensation has been performed for the shown OCT peaks to account for quadratic and third order phase in the experimental setup (see Appendix). The difference in the peak widths can be attributed to slightly different spectral widths of the PDC source due to changes in the waveguide temperature between the measurements. However, the achieved resolution is close to the theoretically estimated resolution of $29.2\,\upmu\mathrm{m}$ for the given PDC spectrum.\\

\begin{figure*}[t!]
	\centering
	\includegraphics[width=\linewidth]{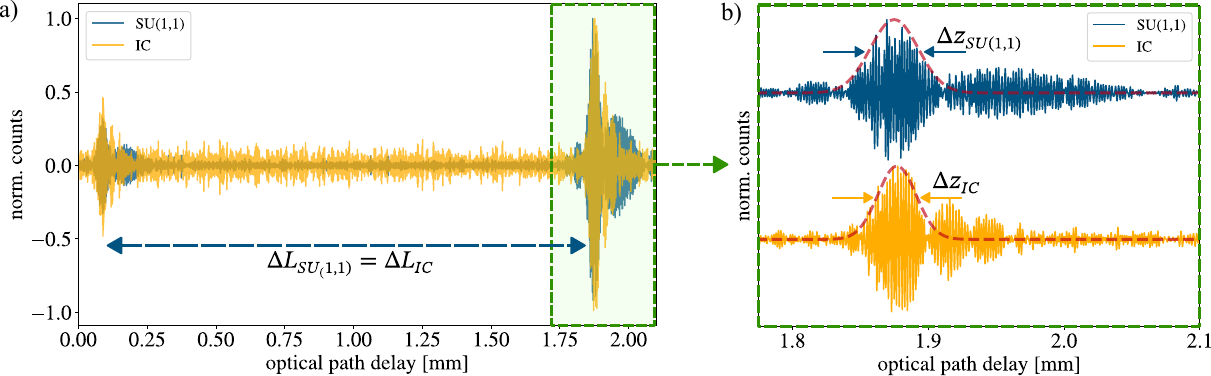}
	\caption{a) TD-OCT signal from the SU(1,1) and IC scheme used to obtain the layer separation$\Delta L_{IC} = (1.80\pm 0.034)\,\mathrm{mm}$ and $\Delta L_{SU(1,1)} = (1.80\pm 0.046)\,\mathrm{mm}$, which is identical for both schemes width . b) A close-up on two peaks in the colored area of reveals a an axial resolution of $\Delta z_{SU(1,1)}=46\,\mathrm{\upmu m}$ and $\Delta z_{IC}= 34\,\mathrm{\upmu m}$ for the SU(1,1) and IC configuration respectively. The count rates in each interferogram are normalised to their maximum. The traces in b) are offset for better visualization and comparison. A Gaussian envelope has been added for the estimation of the axial resolution.} 
	\label{fig:TD_comp}
\end{figure*}

Finally, we measure the same object in TD-OCT at a total pump power of $50\,\mathrm{\upmu W}$ by translating the idler mirror while recording the signal photon counts. Here, we can also extract an OPD of $\Delta L_{IC} = (1.80\pm 0.034)\,\mathrm{mm}$ and $\Delta L_{SU(1,1)} = (1.80\pm 0.046)\,\mathrm{mm}$ in both the IC and the SU(1,1) scheme, as shown in Fig.\,\ref{fig:TD_comp} a). However, we observe a better axial resolution of $\Delta z_{TD,IC} = 34\,\mathrm{\upmu m}$ in the IC scheme in comparison to $\Delta z_{TD,SU(1,1)} = 46\,\mathrm{\upmu m}$ in the SU(1,1) scheme when inspecting the close-up of the interference from one surface in Fig.\,\ref{fig:TD_comp} b). Similar to the FD-OCT, dispersion compensation can be applied to the TD-OCT measurements, which allows to reach an axial resolution of $25\,\mathrm{\upmu m}$ as we show in the Supplemental Material.\\
The IC scheme exhibits a higher axial resolution in TD-OCT compared to the SU(1,1) configuration, because only the idler but not the signal from the first process experiences the dispersion of the waveguide. This increased resolution reduces the need for phase-compensation in post-processing. Thus, the IC scheme is better suited for direct measurements of quantum TD-OCT traces.\\
Comparing the two methods of FD-OCT and TD-OCT we find that they require a similar amount of resources. Meanwhile, TD-OCT can be carried out with less expensive equipment as it only requires a single photon click detector instead of a single photon sensitive spectrograph as the FD-OCT, which makes TD-OCT suitable for cheap, miniaturized quantum sensors.\\  

In order to move further towards practical applications, the SNR of around 300 and 20 in TD-OCT and FD-OCT, respectively, in our proof-of-principle measurements can be increased via longer exposure times or averaging of multiple short measurements. We demonstrate such an increase of SNR due to averaging in the TD-OCT and FD-OCT measurements with undetected photons within our setup in part VII of the Supplemental Material. Although further optimization of the sensor has to be performed for a higher SNR, our experiments show that high SNR values can be reached at lower pump powers compared to bulk implementations of such a sensor which typically operate at a few hundred mW of pump power \cite{Vanselow2020}.

\section{Discussion}

In this work, we provide a first comparison between two experimental configurations for nonlinear interferometers – IC and SU(1,1) – with respect to their suitability for the implementation of practical, integrated quantum sensors for OCT measurements with undetected photons. We have seen that integration comes with new challenges which requires different experimental setups compared to current realizations of nonlinear interferometers based on bulk nonlinear crystals.\\
As a first step, we therefore theoretically described both systems, performed pump gain optimization in both schemes to achieve an optimal gain ratio for a high SNR in the later measurements and compared the underlying mechanisms in both schemes.\\ 
The SU(1,1) scheme, although popular for proof of concept demonstrations based on bulk PDC sources, requires coupling of both photons at the long and the short wavelengths back to the waveguide. Due to spatial multi-modedness at the short wavelength, this configuration is sensitive to minute alignment errors. Furthermore, due to unavoidable losses while coupling back to the waveguide, it is necessary to pump the two PDC sources with different gains, further complicating the experiment. Lastly, in direct TD-OCT measurements, the SU(1,1) system shows a worse performance due to the additional optical path through the dispersive waveguide material.\\ 
In contrast, in the IC configuration, the short wavelength light can be directly coupled to single mode fibres and interfered on an integrated beam splitter, which ensures perfect spatial overlap and high-quality interference. Moreover, light from both PDC processes can be coupled to the fibre with the same efficiency. Therefore, the optimal operation point is reached when both PDC processes are driven with the same pump power. This can easily be achieved with a reflective end-facet coating for the pump field which then reduces the need for a second pump coupling. Slight mismatches in the two gains can be externally compensated for by adjusting the fibre coupling efficiencies. Finally, the resolution in TD-OCT is improved in this scheme compared to SU(1,1) configuration.\\
Considering resource requirements, FD-OCT is the preferred choice if spectrometers are available at the detection wavelength thanks to its potential single-shot performance. For quantum OCT applications, however, we find that the reduced pump power requirements of TD-OCT (two orders of magnitude less than for FD-OCT, see Appendix) combined with the wide availability of cheap single-photon detectors for visible and NIR wavelengths and the, in principle, unlimited axial scanning range renders TD-OCT a better overall choice.\\
Meanwhile, we see that the achieved axial resolution of $28\,\mathrm{\upmu m}$ agrees with the theoretical limit given by the bandwidth of our source and is comparable the axial resolution achieved by bulk systems around $10\,\mathrm{\upmu m}$ \cite{Vanselow2020}. In the future, integrated PDC sources with even larger bandwidths \cite{Roeder2024_source} hold the potential to achieve even higher resolutions.\\
Furthermore, we have seen that our first demonstration of a quantum sensor based on an integrated waveguide source provides a similar SNR compared to already optimized bulk implementations. At the same time, our measurements can be performed at lower required pump powers due to the higher brightness of our source. In the future, further optimization and integration of the presented quantum sensor therefore holds the potential for its use in practical applications.

\section{Appendix}

\subsection*{OCT system benchmarks}

The main benchmark of an OCT system is the axial resolution $\Delta z$. In OCT with undetected photons, this resolution is limited by the spectral bandwidth $\Delta \lambda$ of each of the generated PDC photons from our group-velocity matched, integrated source, c.f. \cite{Pollmann2024}. The maximally achievable axial resolution $\Delta z$ is then given by the coherence length of the source $l_c$ via

\begin{equation}
	\Delta z = l_c = \frac{f}{2}\frac{\lambda_0^2}{\Delta \lambda}.
\end{equation}

In this formula, $c$ is the speed of light in vacuum and $f$ a factor that changes for different shapes of the spectral envelope of the source and can be derived via the Wiener-Khinchin theorem. For our source, which exhibits a rectangular spectrum, this factor is $f=1.2$ \cite{Vanselow2020}. With a spectral bandwidth in the experiment of $\Delta \lambda = 14.2\,\mathrm{nm}$ and central wavelength of $\lambda_0 = 832\,\mathrm{nm}$, this formula predicts an axial resolution of $\Delta z = 29.2\,\mathrm{\upmu m}$.\\
We investigate OCT measurements in two modes of operation, namely time domain (TD-OCT) and Fourier domain (FD-OCT). For TD-OCT, the reference arm needs to be scanned while signal photon counts are recorded to form an interferogram in the temporal domain. Meanwhile, in FD-OCT the spectral interference is observed on a spectrometer for a fixed detuned position of the reference arm and therefore a fixed time delay. Thus, in classical implementations, mostly FD-OCT is used due to higher data acquisition rates and no need for moving parts in the setup. However, these setups are limited in their axial scanning range $\Delta L_z$ due to the limited number of pixels $N$ of the spectrometer used to observe the spectral interference \cite{Wojtkowski2004, Leitgeb2003}, resulting in a maximal axial scanning range of $\Delta L_z = \frac{1}{4} \frac{\lambda_0^2}{\Delta \lambda} N$. For our system at hand with $N=670$ illuminated pixels this scanning range for the round-trip optical path delay is $\Delta L_z = 8.2\,\mathrm{mm}$, which is higher than the tested optical path differences of the used object under test in this work.

\subsection*{Theoretical model}

Here, we describe our model on which we base our OCT measurements with undetected photons for TD-OCT and FD-OCT. We will provide a derivation of the expressions for the spectral and temporal interferograms in both experimental setups in a comprehensive way that is adjusted to our experimental conditions. We describe the IC and SU(1,1) configuration in one expression since the difference in the description of the SU(1,1) and IC geometry reduces to a different phase factor, as we will see later. In order to describe OCT measurements with undetected photons in the two schemes, we start with the two photon component of the PDC state at signal and idler frequencies $\omega_s$ and $\omega_i$ at the output of one source, which can be written as \cite{Uren2009}:

\begin{equation}
	|\Psi \rangle = C \int \mathrm{d}\Delta \omega \cdot f_{PDC}(\Delta \omega)|\bar{\omega}_s -\Delta \omega\rangle \otimes |\bar{\omega}_i + \Delta \omega \rangle
\end{equation}

where C is a constant which is related to the conversion efficiency of the process. The JSA of a single PDC process is then given by 

\begin{equation}
	f_{PDC}(\Delta \omega) = \mathrm{sinc}\left(\frac{\Delta \beta(\Delta \omega)L}{2}\right)\exp(i \cdot \frac{\Delta \beta(\Delta \omega)L}{2}).
\end{equation}

Due to the structure of the bi-layer sample, only a part of the idler generation probability amplitude is directly reflected back to the nonlinear crystal from the first sample layer while the rest acquires a linear phase shift and is reflected back by the second layer. After the interaction with this bi-layer object, our two photon state can be written as:

\begin{equation}
	\begin{aligned}
		|\Psi \rangle_{OCT} = &C \int \mathrm{d} \Delta \omega \cdot f_{PDC}(\Delta \omega)|\omega_s \rangle\\& \otimes \left( r_1 |\omega_i \rangle +  r_2 \exp(i \cdot \frac{2 n_g d  \Delta \omega}{c}) |\omega_i \rangle \right).
	\end{aligned}
\end{equation}

We assume that a maximum of one photon pair per spectral mode is generated at the output of the interferometer. In this low gain regime, we can describe the system in the framework of induced coherence without induced emission \cite{Lemos2014, Wang1991}. We can write the state after the interaction based on the JSA of the two single PDC processes as

\begin{widetext}
	
	\begin{equation}
		\begin{aligned}
			|\Psi \rangle_{OCT} &= C \int \mathrm{d} \Delta \omega \left(f_{PDC}(\Delta \omega) r_1 |\omega_s \rangle|\omega_i \rangle +  f_{PDC}(\Delta \omega) r_2 \exp(i \cdot \frac{2 n_g d  \Delta \omega}{c}) |\omega_s \rangle |\omega_i \rangle \right) \\
			& = C \int \mathrm{d} \Delta \omega \left(f_{PDC}(\Delta \omega) r_1 + f_{PDC}(\Delta \omega) r_2 \exp(i \cdot \frac{2 n_g d  \Delta \omega}{c}) \right) |\omega_s \rangle |\omega_i \rangle . 
		\end{aligned}
	\end{equation}
	
\end{widetext}

Finally, the second source contributes to the output state of the interferometer $|\Psi \rangle_{int}$ as an added term in the form of a single PDC source. Thus, this state can be written as $|\Psi \rangle_{int} = |\Psi \rangle_{OCT} + |\Psi \rangle_{PDC,2}$. We can identify the JSA of this state as a new JSA of the whole interferometer $f_{int}$ and add an adjustable time delay on the idler arm $\tau$. Thus, we end up with

\begin{equation}
	\begin{aligned}
		&f_{int}(\Delta \omega) = f_{PDC}(\Delta \omega) \cdot [r_1 \exp(i \tau \Delta \omega + \Phi_{int}(\Delta \omega))\\& \hspace{1mm} + r_2 \exp(i \tau \Delta \omega + \Phi_{int}(\Delta \omega) + i \cdot 2 n_g d/c \Delta \omega)].
	\end{aligned}
\end{equation}

Here, the phase terms includes internal phases $\Phi_{int}$ from the interferometer itself, e.g., dispersion from the waveguide material. Here, we refer to earlier work for more details on the influence of phases in the SU(1,1) setup \cite{Roeder2024}. Meanwhile, this expression can be applied as well to the IC configuration where only the internal phase $\Phi_{int}$ is different due to the changes for the signal path as depicted in Fig.\,\ref{fig:concept}. From this joint spectral amplitude at the output of the interferometer, we arrive at the expressions for the spectral and temporal interferograms stated above. \\
While this intuitive description is sufficient for our demonstration of OCT measurements with undetected photons based on integrated waveguides, the extraction of information about layered objects under tests in the nonlinear interferometers based on the IC or SU(1,1) geometries have been described in earlier works in more general yet more complex frameworks \cite{Rojas-Santana2021, Valles2018, Machado2020}.

\subsection*{Experimental details}

Both facets of the waveguide are coated with anti-reflection coatings for the pump wavelength, as well as the signal and idler wavelengths. We utilize a folded setup with a single waveguide source to ensure high spectral interference visibility, caused by the two PDC sources being identical by design. Furthermore, the use of a waveguide as a PDC source generally leads to strong frequency correlations in the JSA while spatial correlations between photons do not arise due to their generation in single spatial modes. Thus, all fields in the setup are collinear. For the IC configuration, a 50:50 fibre beam splitter (Thorlabs TW805R5F2) was used to overlap the signal fields from both processes.

\subsection*{Spectral interferograms and visibility extraction}

In the experiment, the spectral interference is recorded at a fixed idler mirror position and for integration times of $1\,\mathrm{s}$ of the spectrograph, if not stated otherwise. The measurements have been performed at pump power levels of around $5\,\mathrm{mW}$, which are required for a significant signal on the single photon sensitive spectrograph (Andor Shamrock SR-500i spectrograph with Newton 970P EMCCD - camera) with a spectral resolution of $30\,\mathrm{GHz}$.\\ 
The value for the interference visibility is obtained by evaluating fringes over the whole signal spectral range. The period of the oscillations is controlled by setting a fixed delay between the signal and idler paths. Two examples for the raw data of the spectral interferograms are presented in Fig.\,S1 in the Supplemental Material. The error bars on the visibility are obtained by standard deviation of the extracted visibility across the spectral bandwidth and while the error bars of the gain ratio are based on the used power meter with a tolerance of $5\,\%$.

\subsection*{Temporal interferograms}

We record temporal interferograms by moving the idler mirror in steps of $200\,\mathrm{nm}$ with a motorized stage (PI M111.1DG) and integrating the counts on a Si-APD (Perkin Elmer Photon Counting Module SPCM-AQR-14FC) and Time-Tagger (Swabian Instruments Time-Tagger 20) for $0.05\,\mathrm{s}$ per step. For these measurements, lower pump powers on the order of $50\,\upmu\mathrm{W}$ are required. Drifts and fluctuations in the recorded count rates have been eliminated by applying a Fourier-filter to the data. This way, the low-frequency noise is filtered out while the information at the frequency of the interference is kept. This frequency is given by half of the central wavelength of the arm in which the stage is moved. In our case this is the wavelength of the idler around $1400\,\mathrm{nm}$.\\   

\subsection*{Resource comparison}

In our current setup, we find that FD-OCT requires pump powers of $5\,\mathrm{mW}$ for sufficient SNR at integration times of $1\,\mathrm{s}$. Meanwhile, the pump power of the TD-OCT measurements has been set to $50\,\mathrm{\upmu m}$ with integration times of $0.05\,\mathrm{s}$ and a total number of $10000$ steps per scan, resulting in a total measurement time of $500\,\mathrm{s}$. The two methods become comparable in measurement time at the same pump power level if we consider the factor of $100$ in the pump powers and assuming the linear connection between pump power and mean photon number which is directly linked to the required integration time.

\subsection*{Numerical dispersion compensation}

The dispersion from the setup has been compensated numerically in the FD-OCT measurements. To this end, a Hilbert transform has been applied to the spectral interferogram. The resulting signal has been multiplied with a quadratic and cubic spectral phase to compensate for spectral phases from the setup. In order to obtain the presented peaks in Fig.\,\ref{fig:FD_comp}, phase factors of $11500\,\mathrm{fs^2}$ and $-45000\,\mathrm{fs^3}$ have been applied in the case of SU(1,1). In the case of IC spectral phases of $600\,\mathrm{fs^2}$ and $-45000\,\mathrm{fs^3}$ have been applied. 

\begin{acknowledgements}
The authors thank Abira Gnanavel and Helen Chrzanowski for helpful discussions. F.R. is member of the Max Planck School of Photonics supported by the German Federal Ministry of Research, Technology and Space (BMFTR), the Max Planck Society, and the Fraunhofer Society. 
We acknowledge financial support from the Federal Ministry of Research, Technology and Space (BMFTR) via the grant agreement no. 13N16352 (E2TPA).
This project has received funding from the European Union’s Horizon Europe research and innovation programme under grant agreement No 101070700 (MIRAQLS). 
\end{acknowledgements}

\bibliographystyle{apsrev4-2}
\bibliography{NIROCTbibo}
\end{document}


\title{Supplementary Material: Towards integrated sensors for optimized OCT with undetected photons}

\author{Franz Roeder}
\email{franz.roeder@uni-paderborn.de}
\affiliation{Paderborn University, Institute for Photonic Quantum Systems (PhoQS), Warburger Str. 100, 33098 Paderborn, Germany}
\affiliation{Paderborn University, Integrated Quantum Optics, Warburger Str. 100, 33098 Paderborn, Germany}
\author{René Pollmann}
\affiliation{Paderborn University, Institute for Photonic Quantum Systems (PhoQS), Warburger Str. 100, 33098 Paderborn, Germany}
\affiliation{Paderborn University, Integrated Quantum Optics, Warburger Str. 100, 33098 Paderborn, Germany}
\author{Viktor Quiring}
\affiliation{Paderborn University, Institute for Photonic Quantum Systems (PhoQS), Warburger Str. 100, 33098 Paderborn, Germany}
\affiliation{Paderborn University, Integrated Quantum Optics, Warburger Str. 100, 33098 Paderborn, Germany}
\author{Christof Eigner}
\affiliation{Paderborn University, Institute for Photonic Quantum Systems (PhoQS), Warburger Str. 100, 33098 Paderborn, Germany}
\author{Benjamin Brecht}
\affiliation{Paderborn University, Institute for Photonic Quantum Systems (PhoQS), Warburger Str. 100, 33098 Paderborn, Germany}
\affiliation{Paderborn University, Integrated Quantum Optics, Warburger Str. 100, 33098 Paderborn, Germany}
\author{Christine Silberhorn}
\affiliation{Paderborn University, Institute for Photonic Quantum Systems (PhoQS), Warburger Str. 100, 33098 Paderborn, Germany}
\affiliation{Paderborn University, Integrated Quantum Optics, Warburger Str. 100, 33098 Paderborn, Germany}

\maketitle
\renewcommand{\figurename}{Fig. S}
\onecolumngrid
\section{Pump gain optimization in the IC configuration}

In this section, we derive the visibilities and optimal pump gain ratios for the induced coherence geometry. To this end, we use the same assumptions as in the derivation of the pump gain optimization in the SU(1,1) geometry in \cite{Santandrea2023}. Based on the illustration of the experimental concept in Fig.\,1 in the main text, the signal photon count rate at one output of the fibre beam splitter is given by

\begin{equation}
	C = T_{800,1} g_1^2 + T_{800,2} g_2^2 + 2\sqrt{T_{800,1}T_{800,2}T_{1400}}g_1g_2 \cos(\Phi)
\end{equation}

where $\Phi$ is the relative phase between the two processes.\\
This leads to a spectral interference visibility $V_{800}$ at the signal wavelength of

\begin{equation}
	V_{800} =\frac{2 \sqrt{T_{800,1}T_{800,2}T_{1400}}g_1g_2}{T_{800,1}g_1^2 + T_{800,2}g_2^2}
\end{equation}

We can see the optimal pump gain ratio for the maximum visibility for measurements with undetected photons with the sample placed in the idler arm is given by

\begin{equation}
	\frac{g_1^2}{g_2^2}_{S} = \frac{T_{800,2}}{T_{800,1}}
\end{equation}

and is therefore independent of any loss introduced by the sample, which is characterized by $T_{1400}$. Meanwhile, the value of the maximally achievable visibility at this optimal pump gain ratio is given by this transmission of the idler arm $T_{1400}$.\\
We see that the pump gain optimization in the IC case is very similar to the one performed in the SU(1,1) setup. However, the conceptual and experimental details allow to achieve an advantage in the implementation of measurements with undetected photons, as we discuss in the main text.

\section{Raw data spectral interferograms}

\begin{figure}[h]
	\centering
	\includegraphics[width=\linewidth]{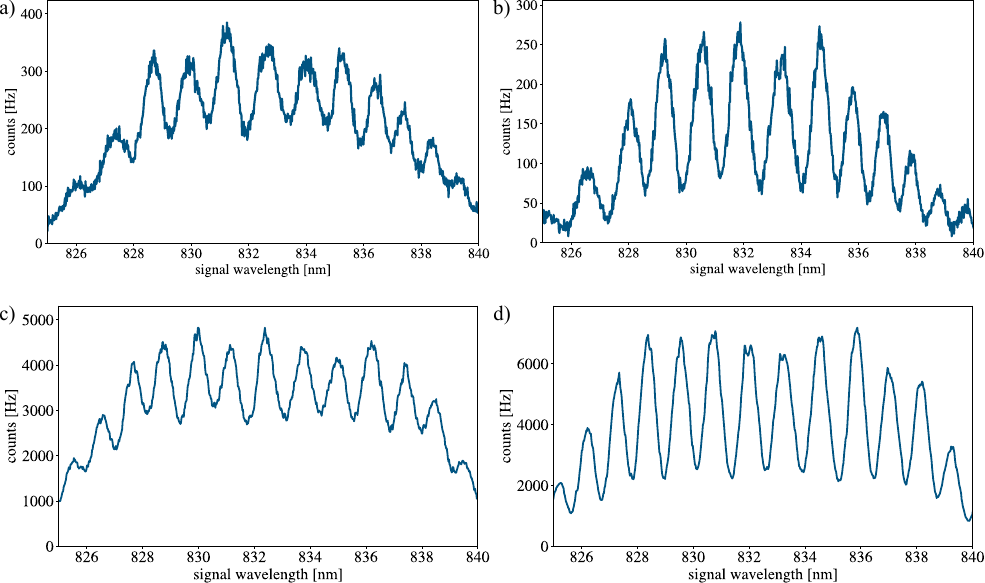}
	\caption{Spectral interferograms for different pump gain ratios and visibilities in the SU(1,1) configuration: a) $g_1^2/g_2^2 = 0.15$, $V=27\,\%$ and b) $g_1^2/g_2^2 = 4$, $V= 59\,\%$ as well as in the IC configuration: c) $g_1^2/g_2^2 = 0.4$, $V=23\,\%$ and d) $g_1^2/g_2^2=0.9$, $V=50\,\%$.}
	\label{Supp:source_spectrum}
\end{figure}

In Fig.\,\ref{Supp:source_spectrum}, we present the raw data of two spectral interferograms with different visibilities due to different pump gain ratios in both experimental configurations. 

The values for the interference visibility are extracted by evaluating the peak prominence of the central fringes between $828\,\mathrm{nm}$ and $838\,\mathrm{nm}$ in each measurement. The error is then given by the standard deviation.

\section{Losses in the signal arm}

\begin{figure}[h]
	\centering
	\includegraphics[width=0.65\linewidth]{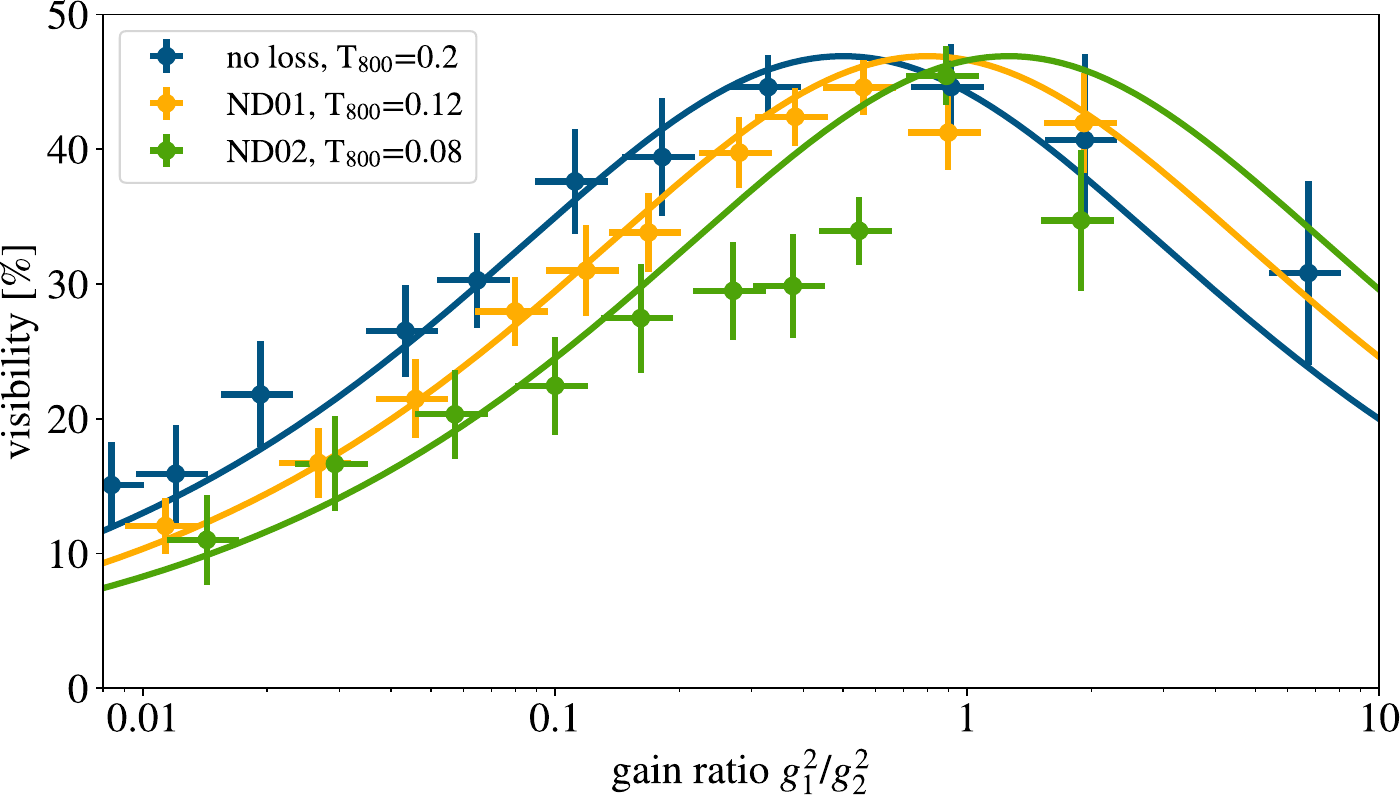}
	\caption{Shift in the optimal pump gain ratio for increasing signal losses by introducing neutral density filters ND01 and ND02 in the signal arm of the interferometer in the SU(1,1) geometry. The solid lines represent the theory based on the given transmission values.}
	\label{Supp:signal_losses}
\end{figure}

Additional losses in the signal arm lead to a shift in the optimal gain ratio, while the maximally achievable visibility stays constant. We show this behaviour in Fig.\,\ref{Supp:signal_losses}. Here, we depict the visibility of spectral interferograms obtained in the SU(1,1) geometry together with the expected theoretical curves. In the measurements, we introduced the neutral density filters ND01 and ND02 into the signal arm after the first PDC process to realise changes in the transmission $T_{800}$.\\
The observed shift in the optimal pump gain ratio allows one to mitigate losses in the system during back-coupling by adjusting the gains.\\
In the IC scheme, losses in one arm would shift the curve in a similar manner. However, since in this scheme the ratio between the transmission values of the two signal paths $T_{800,1}$ and $T_{800,2}$ determine the optimal pump gain ratio, a high visibility can alternatively be achieved by attenuating the opposite signal path in order to stay at the same value for the optimal pump gain ratio.

\section{Sensitivity to idler transmission changes}

\begin{figure}[h!]
	\centering
	\includegraphics[width=0.65\linewidth]{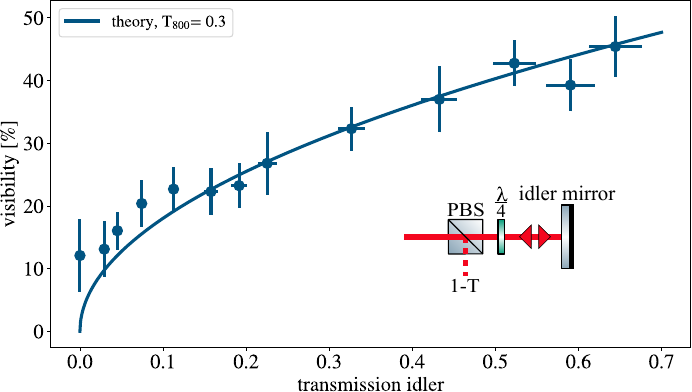}
	\caption{Change in the visibility for different idler transmission values for the optimized gain ratio in the SU(1,1) geometry. The measured values agree well with the theory down to transmissions below $5\%$. The inset shows the variable attenuation in the idler arm with a combination of PBS and QWP in a double pass configuration.}
	\label{fig:variable_losses}
\end{figure}

We studied the limit for the lowest detectable layer reflectivity in our setup by introducing losses in the idler path. For this, we operate at the optimal pump gain ratio and place a polarizing beam splitter and quarter wave plate in a double pass configuration in the idler path. With this, we are able to gradually reduce the idler transmission $T_{1400}$. We evaluate the resulting interference visibility and compare it to the theoretically expected value, as shown in Fig.\,\ref{fig:variable_losses}. This way, we determine the minimal reflectivity of a layer in a potential OCT sample of $5\,\%$ to be detectable. This minimal detectable reflectivity can be reduced further via dedicated data acquisition and utilizing averaging. We performed these measurements only in the SU(1,1) geometry, but expect the same behavior for the IC scheme.

\newpage

\section{Raw data FD-OCT}
\label{app:raw_spectra}

\begin{figure}[h]
	\centering
	\includegraphics[width=\linewidth]{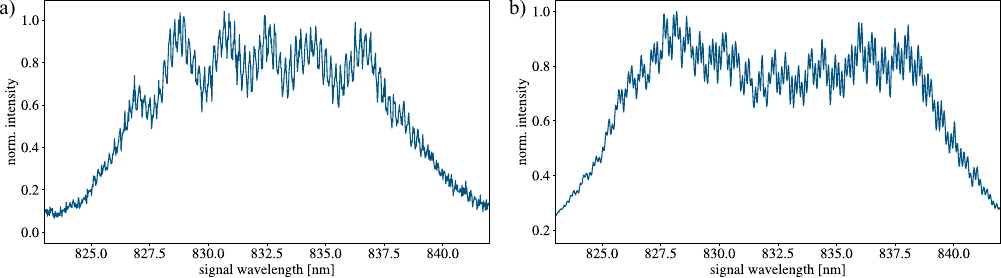}
	\caption{Spectral interferograms with a beating in the fringing period for FD-OCT in a) the SU(1,1) scheme and b) the IC scheme.}
	\label{Supp:OCT_spec_comb}
\end{figure}

Here, we present two sets of raw data of the spectral interferograms from a FD-OCT measurement of a bi-layer object under test in the idler arm in the SU(1,1) and the IC configuration. The measurements are shown in Fig.\,\ref{Supp:OCT_spec_comb}. One can observe a beating of interference fringes, which originate from different temporal delays between and signal and idler photons from the two different reflections off the surfaces of the object under test. The different periods of the oscillations and their visibilities in the two measurements can be attributed to different settings of the delay as well as alignment and parallelism of the layers in the object under test.

\section{Phase compensation FD-OCT}
\label{app:phase_comp}

\begin{figure}[h!]
	\centering
	\includegraphics[width=\linewidth]{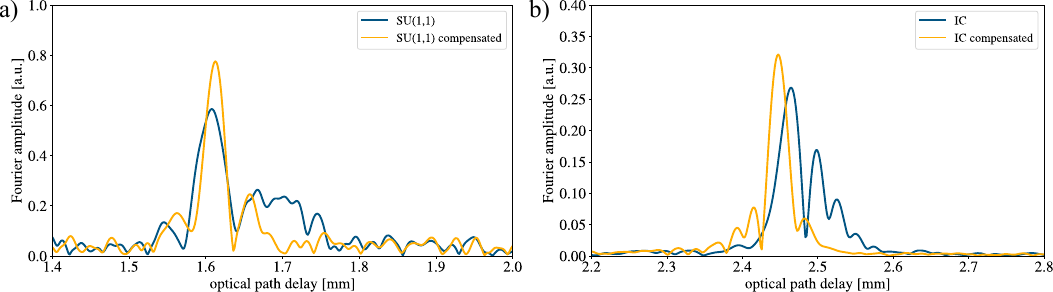}
	\caption{Reduction of the FWHM of the peak from one surface reflection in FD-OCT for a) the SU(1,1) scheme and b) the IC scheme. In both cases the directly retrieved peaks and the dispersion compensated ones are presented.}
	\label{Supp:OCT_comp}
\end{figure}

Spectral phases in the experimental setup, e.g. from the waveguide material or the OCT sample material, lead to a broadening and distortion of the peaks in FD-OCT. The highest axial resolution can be reached by numerically compensating for these phases in post-processing. This is a standard procedure for OCT measurements \cite{Vanselow2020, Wojtkowski2004}. It relies on the application of a Hilbert transform on the measured spectral interferogram, a multiplication with a phase term for compensation, followed by an inverse Hilbert transform. This manipulated spectral interference pattern is finally Fourier-transformed to reveal the depth information in the temporal domain. The applied phase term for compensation is chosen by iteration of the described procedure while minimizing the FWHM of the peak in the temporal domain.\\
A reduction in the peak width in comparison to the originally recorded measurements in the SU(1,1) and IC scheme is shown in Fig.\,\ref{Supp:OCT_comp}. Here, we reach an axial resolution of $28.2\,\mathrm{\mu m}$ and $26.6\,\mathrm{\mu m}$ (FWHM) in both schemes and observe a reduction from initially $86.2\,\mathrm{\mu m}$ and $68.8\,\mathrm{\mu m}$ for the SU(1,1) scheme and IC scheme, respectively. The applied phases for compensation are $11500\,\mathrm{fs}^2$ and $-45000\,\mathrm{fs}^3$ in the SU(1,1) scheme and $600\,\mathrm{fs}^2$ and $-45000\,\mathrm{fs}^3$ in the IC scheme. These phases are dependent on the introduced delay and the spectrometer calibration, as discussed in \cite{Vanselow2020, Wojtkowski2004, Liu2010}.\\
Phase compensation is also possible in TD-OCT and is discussed below on the example of the averaged temporal interferogram.

\section{Effects of averaging}
\label{app:averaging}

\begin{figure}[h]
	\centering
	\includegraphics[width=0.65\linewidth]{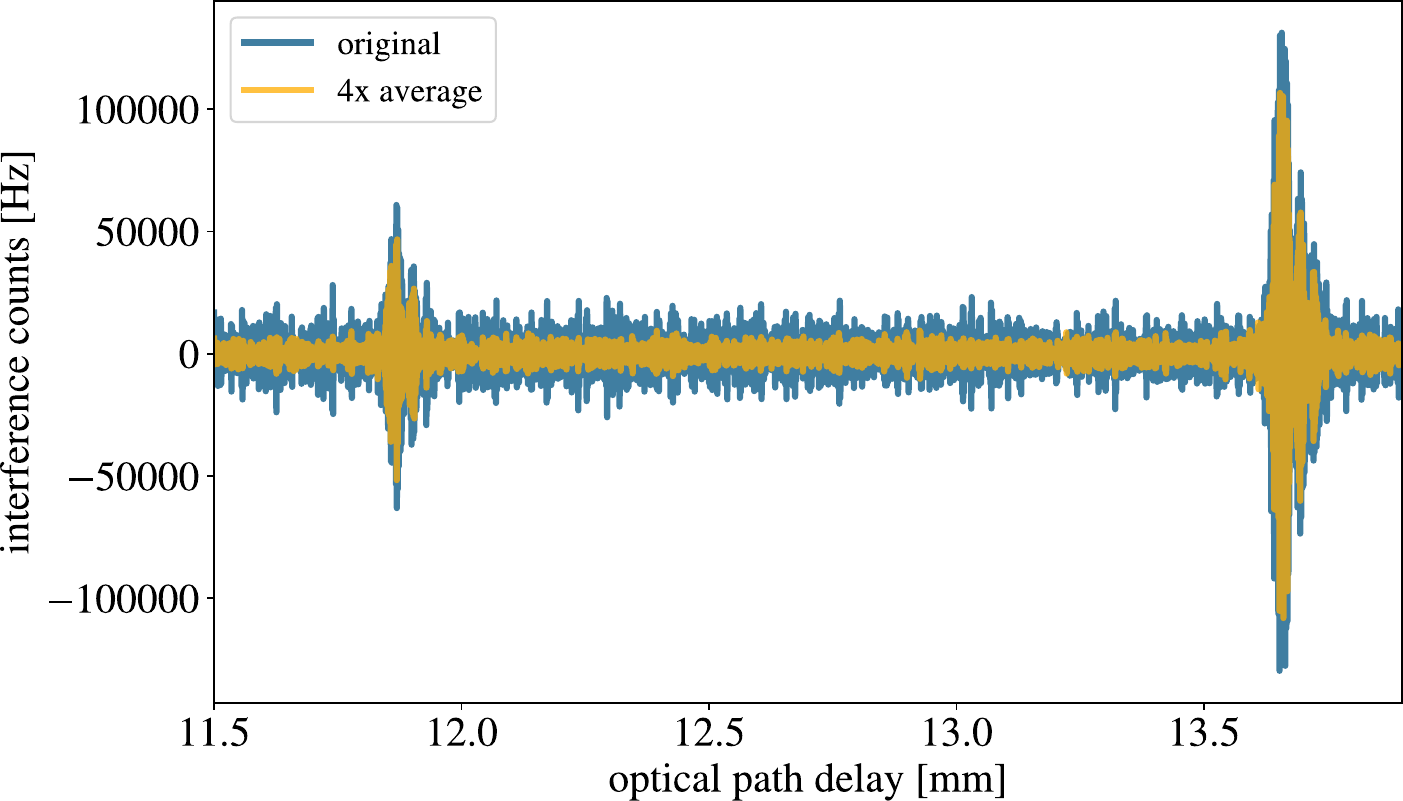}
	\caption{Temporal interferograms for a single acquisition and after averaging over 4 acquisition.}
	\label{App:avg_temp}
\end{figure}

Here, we present the effects of averaging on the signal to noise ratio (SNR) in OCT measurements with undetected photons in our setup. We investigate the effects of averaging on the signal to noise ratio in the TD-OCT and FD-OCT.\\
The improvement due to averaging of multiple temporal interferograms is shown in Fig.\,\ref{App:avg_temp}. We can observe that the noise level between the peaks is reduced while the intensity of the peaks from the different surfaces and their shape remain the same. Due to inaccuracies of the stage position and their repeatability, we overlapped 4 temporal interferograms by shifting them against one measurement serving as a reference in coherent addition as described in \cite{Lindner2021}. For this example, we observe an increase in the SNR from 300 to over 4000. The presented data has been recorded in the IC scheme, however, the technique applies equally to the SU(1,1) setup.

\begin{figure}[h]
	\centering
	\includegraphics[width=0.65\linewidth]{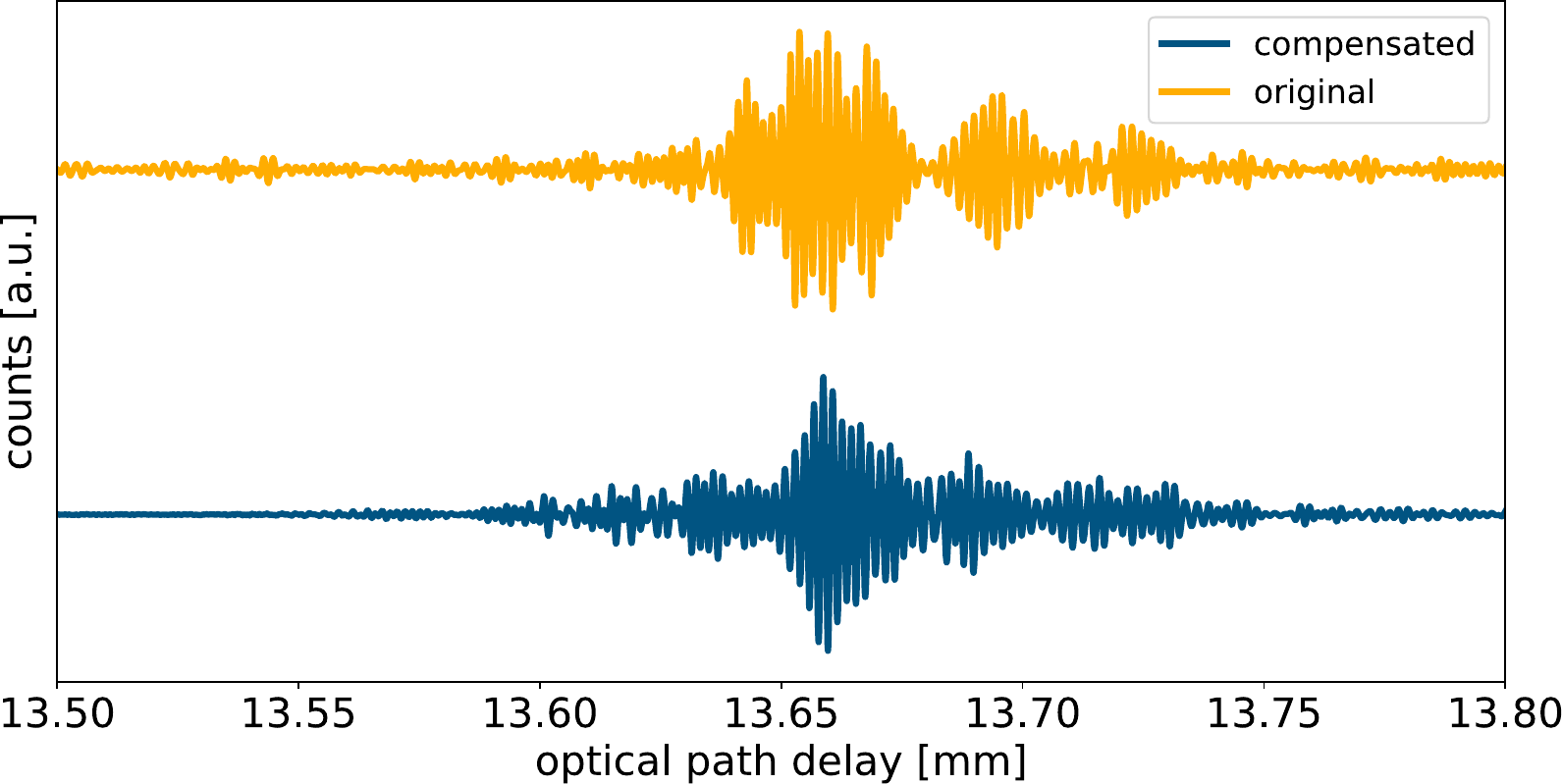}
	\caption{Comparison of the original (yellow) and compensated (blue) temporal interferogram for one speak from the averaged TD-OCT measurement. The data sets are vertically shifted for better visualization.}
	\label{App:comp_temp}
\end{figure}

\begin{figure}[h]
	\centering
	\includegraphics[width=0.65\linewidth]{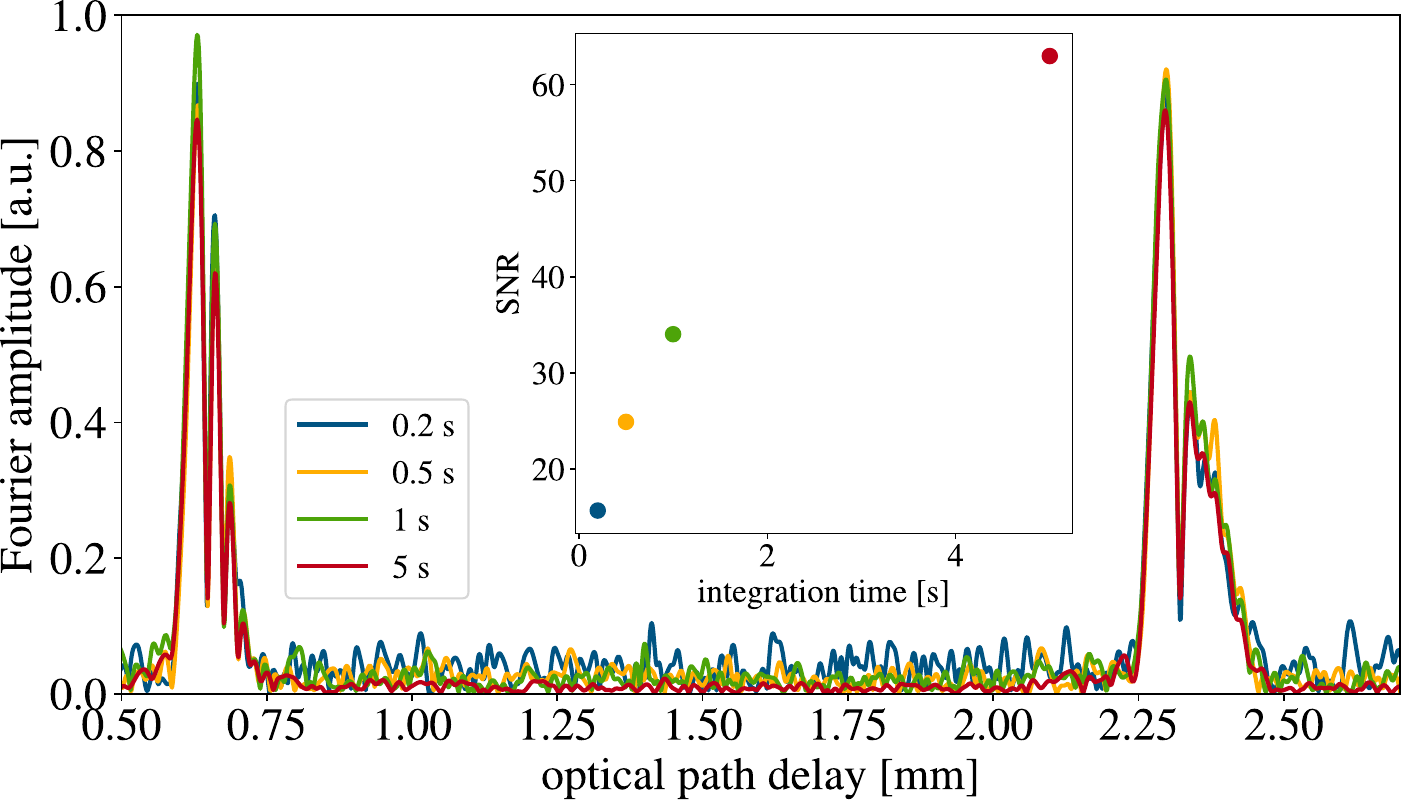}
	\caption{The normalised Fourier transform of a FD-OCT measurement demonstrates a reduction of the noise floor as the integration time increases. Averaging in the spectral interferograms therefore results in an increase in SNR with the integration time of the spectrograph as shown in the inset.}
	\label{App:avg_spec}
\end{figure}

In this averaged temporal interferogram, phase compensation has been performed, which allowed to reduce the width of the main peak from $45\,\mathrm{\upmu m}$ to $25\,\mathrm{\upmu m}$. We note here, that this method requires a good signal to noise ratio and can be carried out even better with more averaged scans. We show the comparison between the original and compensated measurement in Fig.\,\ref{App:comp_temp}. The compensation has been carried out by multiplying the complex spectra in the Fourier-domain by the required quadratic and cubic phase terms to account for the dispersion introduced in the setup. \\
In the FD-OCT, averaging is achieved by increasing the exposure time of the spectrograph and averaging over multiple acquisitions. The resulting change in the FD-OCT intensity and calculated SNR is presented in Fig.\,\ref{App:avg_spec}. We can observe the expected increase in SNR for increasing exposure times. This data has been recorded in the SU(1,1) and applies analogously in the IC scheme.

\newpage
\bibliographystyle{apsrev4-2}
\bibliography{NIROCTbibo}